\documentclass[prd,preprint]{revtex4}
\begin{document}


\title{Late-time Entropy Production from Scalar Decay and Relic
Neutrino Temperature}

\author{Paramita Adhya  and
D. Rai Chaudhuri}
\affiliation{Department of Physics, Presidency College, 86/1, College
Street,\\ Calcutta 700073, India}

\author{Steen Hannestad}
\affiliation{Department of Physics, University of Southern Denmark,
Campusvej 55, 5230 Odense M, Denmark\\
and NORDITA, Blegdamsvej 17, 2100 Copenhagen, Denmark}

\date{13 May 2003}

\bigskip

\begin{abstract}
Entropy production from scalar decay in the era of low
temperatures after neutrino decoupling will change the ratio of the
relic neutrino temperature to the CMB temperature, and,
hence, the value of $N_{eff}$, the effective number of neutrino
species. Such scalar decay is relevant to reheating after thermal
inflation, proposed to dilute massive particles, like the moduli and
the gravitino, featuring in supersymmetric and string theories. The
effect of such entropy production on the relic neutrino temperature
ratio is calculated in a semi-analytic manner, and a recent lower
bound on this ratio, obtained from the WMAP satellite and 2dF galaxy
data, is used to set a lower bound of $\sim 1.5\times 10^{-23}$ Gev
on the scalar decay constant, corresponding to a reheating
temperature of about $3.3$ Mev.
\end{abstract}


\maketitle

\section{Introduction}

The question of neutrino equilibration and
subsequent decoupling
in the early universe has become increasingly
important over the
past few years with the advent of precision
cosmology \cite{cmbnu}.

The canonical textbook result for
neutrino decoupling is that
neutrinos decouple prior to
electron-positron annihilation, leading
to a final neutrino
temperature which is related to the photon
temperature by $r=T_{\nu
0}/T_0 = (4/11)^{1/3}$,
where $T_{\nu 0},T_0$, are,
respectively,
the relic neutrino and CMBR temperatures.

There are quite a
few
factors  which may cause slight departure from the standard
value of
$r=(4/11)^{1/3}$. First, the neutrinos, too, are slightly
heated
during the $e^+e^-$ annihilations \cite{nuheat}, leading to
an
overall increase in neutrino energy density of roughly 1\%.
Finite
temperature
QED effects lead to an additional slight heating of the
neutrinos
\cite{qed}. However, all of these results are within the
standard
model and lead to an overall increase in neutrino energy
density of a
little more than 1\%, an effect which should be included in
practical
calculations.

At present this small effect is not detectable, but
it has
been
estimated that future high-precision measurements of the CMB
anisotropy could reach the required level of
sensitivity
\cite{cmbnu}.

In the present paper we discuss an
additional factor which
has effect on the value of $r$.
Various long-lived,
massive fields like the gravitino, the
Polonyi, the moduli, and
the dilaton,
which figure in supersymmetric
and string theory models
\cite{massive}, pose
cosmological problems,
because their decay must affect $\eta$ and
nucleosynthesis
\cite{intro}. To dilute them away, the proposal
of thermal
inflation
\cite{lyth} has been mooted. A scalar field, the flaton,
is used
to
generate inflation at late times, typically at temperatures of
about
$10^7$ Gev. The inflation stops when the temperature falls to
the
flaton mass $\sim $ $10^3$ Gev. Such a particle will go on
decaying
into the era of Mev-scale temperatures, and affect
nucleosynthesis
and the CMBR.
\par The effect on neutrino
decoupling and
nucleosynthesis has been
studied
\cite{kawa,moroi,kohli,pa}. The studies do not depend
materially
on the
details of flaton phenomenology, and apply to the
Mev-scale
decay of any
scalar $\phi$.
\par In this manuscript, we are trying to see how the
outpouring of
entropy, as $\phi$ decays, heats up the $e^{-},e^{+},
\gamma $
plasma,
and affects the ratio $r$. The assumption of current
models is
that
the $\phi$ should not decay into neutrinos directly \cite{kawa}.
So,
the decay of $\phi$ operates in a direction opposite to
neutrino
heating, and may offset conclusions drawn on the basis of
neutrino
heating. The effect of changing the lower bound on the
scalar
decay
constant $\Gamma$ should be an interesting input in
current
calculations of nuclear abundances and CMBR anisotropies.

Ref.
\cite{kawa} deals directly with the neutrino distribution
function
and
proceeds by solving the Boltzmann equation numerically.
We try,
on the other
hand, to adhere as far as possible to the
macroscopic entities like
temperature and entropy so as to proceed
analytically and keep the
physical
processes transparent.

Of course, this means that our
calculation does not
reach the level
of precision of a numerical solution to the Boltzmann
equation.
However, it does provide a reasonable bound on the
involved
parameters.\par
The plan of the paper is as follows. Section 1 is this Introduction.
In section 2, the entropy production due to scalar decay, since
neutrino decoupling, is estimated, and the basic relation whereby it
affects the relic neutrino temperature ratio is set out. In section
3, the values of this ratio for different values of the scalar
decay constant are calculated. Section 4 uses a lower bound on the
relic neutrino temperature ratio from the WMAP satellite data, the
2dF galaxy
survey and other related data to find a lower bound on the
scalar decay constant, and discusses the results.



\section{Calculation
of the Entropy Production}

Kawasaki et al \cite{kawa} find that the
electron and
muon/taon neutrino distribution functions show a
deficit from
the
thermalised F.D. distribution as the reheating temperature for
the
scalar (and, hence, its decay constant) falls.  This leads to
a
decrease in the $\nu_e$ and $\nu_{\mu,\tau}$ energy densities
with
concomitant effect on the weak interaction rates and
freeze-out
times.
The threshold reheating temperature $T_R$ is around $7$ Mev,
below which the
authors find the effective number of neutrino types
$N_{eff}$ falls below
the value 3.
\par So, even before decoupling, the neutrinos cannot be
assigned the
photon temperature. Can they be assigned a
temperature at all,
in
particular a lower temperature which will approximately reproduce
the
decreased distribution? Strictly speaking, the shape of
the
distribution does not permit this. Ref. \cite{kawa}
defines
$\bar{T}_{\nu}=[2\pi^2n_{\nu}/(3\zeta(3))]^{(1/3)}$
and
$\bar{R_E}=(\rho_{\nu}/n_{\nu})/(3.151\bar{T}_{\nu})$, and
finds that
$\bar{R_E}$
takes values $1.00, 1.03, 1.50$ for $T_R$ values $10,3,1$
Mev,
respectively, while a F.D. distribution should give
$\bar{R_E}=1$. On
the
other hand, these results indicate that unless the decay
constant
$\Gamma$ is much smaller than that required to save
standard BBN,
the
assumption of a neutrino temperature $T_{\nu }$ a little smaller
than
the photon temperature $T$, even before decoupling, is not a bad
one,
especially if one is interested only in locating the parameter
region
where $\Gamma$ does not affect the standard picture.
After
decoupling, the neutrinos are no longer affected by the effects
of
scalar decay, and the momentum of the neutrinos, assumed here to
be
massless, redshifts as $1/a$, so that one can consider the
usual
relic
neutrino temperature-like parameter which redshifts as $1/a$
from the
neutrino decoupling temperature $T_{\nu i}$. \par However,
the main
difference between $T_{\nu }$ and $T$ must arise from the
$e^-,e^+$
annihilations, after the decoupling era, and we are
interested in
seeing how
this is affected by scalar decay. In what
follows, we will neglect
the
difference between $T_{\nu i}$ and
$T_i$, just at the epoch of
decoupling
$t=t_i$, while recognising that the decoupling temperature
remains
uncertain
to some extent. To take this into account, we will consider
a
decoupling
temperature range of $1-3$ Mev.

We have to deal
with
four epochs. $i$ is
an epoch just after electron neutrino,
$\nu$, decoupling, when we
take
$T_{\nu i}= T_{i}$, and

\begin{equation}
g^*_{Si}\approx g^*_i =
11/2
\label{eq:gi}.
\end{equation}
$f$ is an epoch after
$e^{-},e^{+}$
annihilation, which we can take to be the present,
with
neutrino
temperature $T_{\nu 0}$ and photon temperature $T_0$,
and
\begin{equation}g^*_{Sf}= g^*_f = 2\label{eq:gf}.
\end{equation} $a$
is
the epoch of  $e^{-},e^{+}$ annihilation, when the photon
temperature is
$T_a$, and, although in this epoch, the effective
number of
degrees of
freedom is actually changing, we will take
\begin{equation}g^*_{Sa}
\approx
g^*_{a} \approx g^*_{Si}\approx
g^*_i \approx
11/2\label{eq:ga}.\end{equation}\par A basic assumption is
that at some earlier
era, $\rho_{\phi}$ dominated the energy density.
We will need a fiducial era
for $\phi$, when this $\phi$-domination
of the universe ends and
radiation
domination begins. This, we assume,
happens sometime before
decoupling of
the three families of
neutrinos, at a temperature $T_E$, with
$g^*_{SE}\approx g^*_E =
43/4$.
\par The basic relation
is
\begin{equation}g^*_{Si}(2\pi ^2/45)a_i^3T_i^3 +
\Delta S =
g^*_{Sf}(2\pi
^2/45)a_f^3T_f^3.\label{eq:basic}\end{equation}
$\Delta
S $ is the entropy
poured into the plasma between $T_i$ and $T_f$, in
accordance
with the
assumption that $\phi$-decay into neutrinos is
not allowed.
It is to be
calculated from
\begin{equation}dS=-d(a^3\rho_{\phi})/T,
\end{equation}
$\rho_{\phi}$
being the scalar
energy density at radiation temperature T,
radiation
including all massless particles in equilibrium with
photons.
If
$\rho_{R}$ is the radiation density, temperature is defined
from
\begin{equation}
\rho_{R}=(\pi^2/30)g^*T^4\label{eq:rhoR}.
\end{equation}
Taking the
scalar decay rate constant to be $\Gamma$
\cite{three}, the equation for the
evolution of the scalar energy
density
is\begin{equation}\frac{\partial}{\partial
t}
\rho_{\phi}+3H\rho_{\phi}= -
\Gamma
\rho_{\phi}.\label{eq:phidot}\end{equation}
Defining
$\Phi=a^3\rho_{\phi}$   and  $R=a^4\rho_R$, where $a$ is
the
scale factor, (\ref{eq:phidot}) becomes\begin{equation}
\dot{\Phi}=
-
\Gamma
\Phi,\label{eq:Phidot}\end{equation}with the
solution \cite{scherrer}
\begin{equation}\Phi=\Phi_E
e^{-\Gamma
(t-t_E)},
\end{equation} where
$\Phi_E$ is the
value of
$\Phi$ at $t=t_E .$ So,\begin{equation}\dot{S}=(\Gamma
/T)\Phi_E
e^{-\Gamma (t-t_E)}\label{eq:Sdot}.\end{equation}In
(\ref{eq:Sdot}),
the exponential will dominate at the epochs
$i$ and $f$, and in between,
when
$\rho_{R}\gg \rho_{\phi}.$ So, in the pre-exponential, we
may
approximate T from (\ref{eq:rhoR}) using the full radiation
domination
equations $H=1/(2t)$ and $H^2=8\pi\rho_{R}/(3M_{Pl}^2).
$
This gives the
usual relation
\begin{equation}1/T=\alpha\sqrt{ t},
\mbox{with}\:
\alpha^2=2.7215\times
10^{-19}g^{*\frac{1}{2}}Gev^{-1}.
\label{eq:alpha}
\end{equation}Introducing
the variable
\begin{equation}y=\Gamma t=\Gamma /(\alpha^2
T^2)\label{eq:y},
\end{equation} we get \begin{eqnarray}
\Delta S & = &
\Gamma \Phi_E
e^{\Gamma t_E}\int _i^f\alpha
\sqrt{ t}e^{-\Gamma t}dt\nonumber
\\ & =
&  (1/ \sqrt{\Gamma}) \Phi_E e^{\Gamma
t_E}\int
_i^f\alpha\sqrt{y}e^{-y}dy\nonumber \\ & = & (1/
\sqrt{\Gamma})
\Phi_E e^{\Gamma
t_E}\alpha_i[\int_i^a\sqrt{y}e^{-y}dy\nonumber \\ &
+ &
(\frac{4}{11})^\frac{1}{4}(\int
_a^f\sqrt{y}e^{-y}dy)]
\label{eq:DeltaS},
\end{eqnarray} where
$\alpha_i$ is the value of $\alpha$ with
$g^*=g^*_i$ and
$\alpha_f/\alpha_i=(\frac{4}{11})^\frac{1}{4}.$
We have
assumed $g^*$
to be $g^*_i$ from $t_i$ to $t_a$ and
$g^*_f$ from $t_a$
to
$t_f$.

\subsection{Estimate of $\Phi_E$}
Because
there are as yet no
firm phenomenological values for
$\Phi_E$, we have to make an
estimate. In
the expression for
$\dot{S}$ given in (\ref{eq:Sdot}), the
exponential
dominates at the
epochs we are interested in. So, we estimate
$\Phi_E$ in
the
pre-exponential just to an order of magnitude, defining the
epoch
$E$
by taking $\rho_{R}=\rho_{\phi}$ at $t=t_E$. Just for
estimating
$\Phi_E$ in the pre-exponential, we use the crude
approximation
$a\propto 1/T$, although, actually, in the
presence of
entropy
generation there is departure from this type of
evolution, and,
in
fact, in the regime of full $\Phi$ domination, when
$\rho_{\phi}\gg
\rho_R$, $T\propto a^{-\frac{3}{8}}$
\cite{scherrer,kolb}.
In the
regime of $\rho_R\gg
\rho_{\phi}$, of course, we expect a closer fit
to $a\propto
1/T$.
\par With this approximation, in
the
pre-exponential,\begin{eqnarray}\Phi_E &=&
a_E^3\rho_{\phi
E}\nonumber
\\ & =& a_E^3\rho_{RE}\nonumber \\ & =
& (\pi^2/30)g^*_E
a_E^3T_E^4\nonumber
\\ &
=&
(\pi^2/30)g^*_Ea_i^3T_i^3T_E\label{eq:PhiE}.
\end{eqnarray}

\subsection{Estimate of $T_E$}

$T_E$ is the temperature when
$\Phi$-domination passes into
$R$-domination,
i.e. when $H^2$ passes
from
$\frac{8\pi}{3M_{Pl}^2}\rho_{\phi}$ to
$\frac{8\pi}{3M_{Pl}^2}\rho_R$.
So, we try to find approximations for
$H^2$ closer to the epoch
$\rho_{\phi}=\rho_R$, and on either side of
it.
We first consider the era
before this, i.e. the era of incomplete
$\Phi$-domination. From\[
\frac{\partial}{\partial
t}[a^3(\rho_{\phi}+\rho_R)]
+p_R\frac{\partial}{\partial
t}a^3=0,\]and
(\ref{eq:phidot}),(\ref{eq:Phidot}), one
obtains\begin{equation}
\dot{R}=a\Gamma\Phi
\label{eq:Rdot}.\end{equation}The
Friedmann
equation can be written as\begin{equation}H^2
=
\frac{8\pi\Phi}{3M_{Pl}^2a^3}(1+\frac{R}{\Phi
a})
\label{eq:FriPhi}\end{equation}
In the era of incomplete
$\Phi$-domination, the term
$\frac{R}{\Phi a}$ on
the RHS of
(\ref{eq:FriPhi}) is a correction term, and may
be evaluated to
the
approximation\begin{equation}R-R_I=
\frac{dR}{da}|_{a_I}(a-a_I),
\label{eq:Rapprox}
\end{equation}where
$t_I$
refers to some initial epoch such that $a\gg a_I$.
Also, if it
is supposed
that the scalar decay produces sufficiently copious
radiation,
$R_I\ll R$.
As a correction term is being dealt with,
these
approximations should not
cause much deviation from the actual
evolution. Then, for
$t$ sufficiently
later than $t_I$, but within
the regime under consideration,
one may write,
in the correction term
on the RHS
of
(\ref{eq:FriPhi}),\begin{equation}R\approx
\frac{\Gamma}{H}\Phi
a\label{eq:RPhi},\end{equation}using
(\ref{eq:Rdot}) and
(\ref{eq:Rapprox}).  As we are hoping to
find only an
order of
magnitude estimate of a lower bound on
$\Gamma$, (\ref{eq:RPhi})
is
not a bad approximation. Thus, if we go even to very early
times in
this
era, when $\Phi=\Phi_I\approx $ constant\cite{kolb}, integration
of
(\ref{eq:Rdot}) leads to\[ R\approx
\frac{2}{5}\frac{\Gamma}{H}\Phi a
,\]using the approximations $a\sim
t^{\frac{2}{3}}, \: H=
\frac{2}{3t}$. \\
So, introducing a new
evolution variable
$x=\Gamma/H$, we write, for use in
the correction
term on the RHS of
(\ref{eq:FriPhi}),\begin{equation}x=\Gamma/H\approx
R/\Phi
a=\rho_R/\rho_{\phi}\label{eq:defx}.\end{equation} Next,
we
consider the era
of interest to us when $\Phi a\ll R$, such that
$\Phi a/R$ cannot be
neglected, but its higher powers can. In this
era of incomplete radiation
domination, the Friedmann equation is put
in the
form
\begin{equation}H^2=\frac{8\pi
R}{3M_{Pl}^2a^4}(1+\frac{\Phi
a}{R})\label{eq:FriR}.\end{equation}
If, well
into this epoch, the
correction term $\Phi a/R$
on the RHS of (\ref{eq:FriR})
is
neglected, the full
radiation domination relations are
found:\begin{eqnarray}H
&=&\frac{1}{2t},\mbox{ and }\nonumber
\\
a&=&At^{\frac{1}{2}},\label{eq:rdom}\end{eqnarray} $A$ being
a
constant.\par (\ref{eq:Phidot}) has, as solution,
a falling
exponential
in $t$, viz. $\Phi\sim e^{-\Gamma t}$. Instead of taking
the falling
exponential in $t$ directly, a suitable approximation to
the correction term
on the RHS of (\ref{eq:FriR}) is first worked
out. Let $t_0$ be a
sufficiently late epoch, when $\Phi=\Phi_0\approx
0$.
Then, for use only in
the correction term on the RHS of (\ref{eq:FriR}), one
takes
\begin{eqnarray} \Phi -
\Phi_0=
\tilde{\Phi}(\frac{1}{t})-\tilde{\Phi}(\frac{1}{t_0})=
\frac{d\tilde{\Phi}}{d\frac{1}{t}}|_{t_0}(\frac{1}{t}-
\frac{1}{t_0}).
\nonumber\end{eqnarray}
Neglecting $\Phi_0,1/t_0$ compared to
$\Phi,1/t$,
respectively,
an
approximation\begin{equation}
\Phi\approx\frac{B}{t},
\label{eq:Phiapprox}\end{equation}will
be used
only in the correction term on the RHS of (\ref{eq:FriR}),
i.e. in
the correction term, the falling
exponential will be
approximated by a
rectangular hyperbola. B is a
constant.  A similar approximation is
considered for $R$.  It ought
to be mentioned that
$R$ refers to the total
radiation present, and
not only to that produced by decay.
However, the
change in $R$ is due
to $\phi$ decay and consequent entropy
production. In
the absence of
this decay, $\dot{R}=0$.\par Using (\ref{eq:rdom})
and
(\ref{eq:Phiapprox}) in (\ref{eq:Rdot}), and, integrating,
one
obtains,
for use only in the correction term on the RHS of
(\ref{eq:FriR}),\[R-R_E\approx
2AB\Gamma(t^{\frac{1}{2}}
-t_E^{\frac{1}{2}}),\]an
approximation which
corresponds to (\ref{eq:Rapprox}), because of
$a\approx
At^{\frac{1}{2}}$. If $t_E$ is sufficiently early
compared to
$t$,
though within the regime under consideration, and there
is
sufficiently copious radiation production since $t_E$,
it is
sufficient
to take \[R\approx 2AB\Gamma t^{\frac{1}{2}}\] in the
correction term on the
RHS of (\ref{eq:FriR}). This relation, together
with (\ref{eq:rdom}) and
(\ref{eq:Phiapprox}), are now
used to give, in the correction
term on the RHS
of (\ref{eq:FriR}),
\begin{eqnarray}
x &=
&\frac{\Gamma}{H}\nonumber \\ &
\approx & \frac{R}{\Phi
a}\label{eq:defx1},
\end{eqnarray}once again, as
in
(\ref{eq:defx}).\par Introducing the variable
$x$ in (\ref{eq:FriR}),
we
get (\ref{eq:FriR}) in the
form
\begin{equation}\frac{\Gamma^2}{x^2}=
\frac{8\pi
R}{3M_{Pl}^2a^4}(1+\frac{1}{x}).
\label{eq:Hrad}\end{equation}This
is a good equation
for incomplete radiation domination when $x\gg 1$.
However, we will approximate $T_E$ in (\ref{eq:PhiE}), i.e. in the
pre-exponential of (\ref{eq:Sdot}), by putting
$x=x_E\approx 1$ for $t=t_E$
in (\ref{eq:Hrad}).

Now,
(\ref{eq:defx}) and
(\ref{eq:defx1}) signify
that our approximations
are equivalent to taking
$\rho_{\Phi}=\rho_R$ when
$\Gamma =H$, to an
order of magnitude, in the correction
term in $H^2$, and
this is our
way of bypassing lack of knowledge about the
initial value of
$\Phi$.
This approximation has
been explained in the preceding parts of
this
section.  It differs from a common approach to the problem where
decay
is supposed to occur at $t=1/\Gamma $ and $\Phi$ is put
equal to
the
scalar mass at this epoch. Here, we allow the scalar to decay
over
time, but the price of bypassing knowledge about scalar
energy
density
or mass is paid by the approximation inherent in
(\ref{eq:defx}) and
(\ref{eq:defx1}) and our way of estimating
$T_E$. The result for $T_E$ is
\begin{equation}T_E\approx 2^{\frac{1}{4}}\sqrt{\Gamma}/\alpha_E
\label{eq:TE}.\end{equation}If
we use (\ref{eq:defx})
in the incomplete
$\phi$-domination case when
$x\ll 1$, equation
(\ref{eq:FriPhi})
becomes\begin{equation}
\frac{\Gamma^2}{x^2}=\frac{8\pi\Phi}{3M_{Pl}^2a^3}(1+x).
\end{equation}This
gives the same value for $T_E$ if this equation
is extrapolated to
$x=x_E\approx 1$ for $t=t_E.$  Putting the value
of $T_E$ from
(\ref{eq:TE}) in (\ref{eq:PhiE}), (\ref{eq:DeltaS})
becomes
\begin{eqnarray}\Delta S &=&
2^{\frac{1}{4}}(\pi^2/30)g^*_Ea_i^3T_i^3
e^{\frac{1}{2}}(\alpha_i/\alpha_E)[\int_i^a\sqrt{y}e^{-y}dy\nonumber
\\ & +
&(\frac{4}{11})^\frac{1}{4}(\int_a^f\sqrt{y}e^{-y}dy)]
\label{eq:DeltaS1},
\end{eqnarray} Now, in
(\ref{eq:basic}),
(\ref{eq:DeltaS1}),
\begin{equation}a_i^3T_i^3=a_f^3T_{\nu0}^3
\label{eq:Tnu},
\end{equation}because
the temperature of the
decoupled neutrinos red-shifts as $1/a$.  Using
(\ref{eq:Tnu}) and
(\ref{eq:DeltaS1}), we
get\begin{equation}r^3+2.431r^3[ \int_i^a\sqrt{y}e^{-y}dy+
(\frac{4}{11})^\frac{1}{4}(\int_a^f\sqrt{y}e^{-y}dy)]=4
/11\label{eq:basic1},\end{equation}where we
have put
$r=T_{\nu0}/T_0,$ and
taken
$\alpha_i/ \alpha_E=(g^*_i/g^*_E)^{\frac{1}{4}}=
(22/43)^{\frac{1}{4}}.$


\section{Results}

From
(\ref{eq:y}),(\ref{eq:alpha}),(\ref{eq:gi}),(\ref{eq:gf}) and
(\ref{eq:ga}), we
find
\begin{equation}y_i=\frac{1.56675\Gamma_0}{(T_i/Mev)^2},y_f=
\frac{2.60\Gamma_0}{(T_f/Mev)^2},y_a=\frac{1.56675\Gamma_0}
{(T_a/Mev)^2}.\end{equation}$T_i$ is a few Mev. Here, we
calculate
results
for $T_i=1,2,3$ Mev. $T_f$ being $<<1eV$, $y_f$ can be put
$
\approx \infty
$ because of the nature of the incomplete Gamma
function
$\Gamma(1.5,x)$,
where
\begin{equation}\int_{y_1}^{y_2}\sqrt{y}e^{-y}dy=
\Gamma(1.5,y_1)-
\Gamma(1.5,y_2).\nonumber\end{equation}Also,
the properties
of the incomplete Gamma
function $\Gamma(1.5,x)$
indicate
that the results should be insensitive to the precise value
of
$T_a$ in the range $0.3<T_a/Mev<0.5$. We have checked
this
numerically for $T_a/Mev=0.3,0.4,0.5$. The values of the
ratio $r$
for different values of $\Gamma_0$ are displayed in Table
1.
\begin{table} \begin{center}
\begin{tabular} {|l|r |l|r |l|r|}\hline
\multicolumn{2}{|c|}{$T_i=1Mev$}
&
\multicolumn{2}{|c|}{$T_i=2Mev$} &
\multicolumn{2}{|c|}{$T_i=3Mev$}
\\ \hline
\multicolumn{1}{|c|}{$\Gamma_0$} & $r$ & \multicolumn{1}{|c|}
{$\Gamma_0$} & $r$ &\multicolumn{1}{|c|} {$\Gamma_0$} & $r$ \\
\hline
1.0&0.587&4&0.587&10&0.599
\\ \hline
1.2&0.608&5&0.612&12&0.620
\\ \hline 1.4&0.626&6&0.635&13&0.630
\\
\hline 1.5&0.635&7&0.654&14&0.639
\\ \hline
1.6&0.643&8&0.669&15&0.648
\\ \hline 1.8&0.657&10&0.690&16&0.656
\\
\hline 2.0&0.669&12&0.702&20&0.680
\\ \hline
3.0&0.702&14&0.708&24&0.695
\\ \hline 4.0&0.711&20&0.713&28&0.703
\\
\hline 10&0.714&25&0.714&32&0.708
\\
\hline 25&0.714&50&0.714&50&0.713
\\ \hline -&-&-&-&75&0.714 \\
\hline -&-&-&-&100&0.714
\\ \hline
\end{tabular}
\caption{Relic
Neutrino Temperature Ratio $r$ for different values of the
Scalar
Decay
Parameter
$\Gamma_0=\Gamma/(10^{-24}Gev)$}
\end{center}
\end{table}



\section{Conclusions}

A semi-analytical
method for calculating the change in the relic neutrino
temperature ratio
and, hence, in
the effective number of
neutrino species, resulting from
reheating at very low temperatures,
has been presented.

While this
calculation obviously does not have the accuracy of a
full numerical
solution of the Boltzmann equation, it does have
the merit of being
transparent and easily reproducible.

In Ref.~\cite{kawa}, a lower
bound on
the reheating temperature
was calculated from a consideration of
the impact
of incomplete
neutrino equilibration on big bang nucleosynthesis.
While
this bound is powerful, it has the problem that it is flavour
dependent in
the sense that incomplete electron neutrino equilibration
has a direct
impact on the $n-p$ conversion rate.

Here we use, instead,
a bound which relies on energy density only.
Recent results
\cite{steen} show that an
overall best fit for the WMAP $T$ and
$TE$ data, combined with the
Wang {\it et al} compilation of CMB data
\cite{wang}, the
2dFGRS data, the HST
key project data on $H_0$, and the SNI-a
data on
$\Omega_m$, give
bounds on $N_{eff}$ :
\begin{equation}1.9<N_{eff}<7.0 \, (95 \%
\,
\mbox{ confidence }).\end{equation} This corresponds to a
lower bound
on $r$ :
\begin{equation}r>0.637.\label{eq:rbound}\end{equation}\par

Table I shows that the change in $r$ due to $\Gamma$ is sensitive
to
$T_i$,
i.e. the neutrino decoupling temperature.  The lower bound on
$\Gamma$ for a
lower bound of $0.637$ on the relic neutrino
temperature ratio,
corresponding to (\ref {eq:rbound}), is found to
be between
$1$ and $2\times 10^{-24}Gev$, $5$ and $7
\times
10^{-24}Gev$, $12$ and $15 \times 10^{-24}Gev$, respectively,
for
$T_i= 1,2,3$ Mev.  We conclude that to ensure that the relic
neutrino
temperature ratio $r=T_{\nu 0}/T_0$ does not exceed the
lower
bound
of $0.637$\cite{steen}, the scalar decay constant $\Gamma$ must
be
greater than about $15\times 10^{-24} Gev$. Taking the
definition of
the
reheating temperature in \cite{kawa}, this corresponds to a
reheating
temperature of $3.3$ MeV, which is comparable to the
result found in
Ref.~\cite{kawa}.
In their calculation, $N_{eff} = 1.9$
corresponds roughly
to $T_R
\simeq 2.2$ MeV.

This means that even though the
approximations
we use are rough,
the end result is very similar to what is
found from the
full
numerical solution. The reason is that we are only interested
in
energy density, not in the underlying neutrino
distribution
function. On
the other hand, for BBN purposes the distribution function
of
$\nu_e$ is
very important, because high energy neutrinos have
more weight in the $n-p$
conversion processes.
It should also be noted that
the lower bound on
$T_R$ of 0.7 MeV
found in Ref.~\cite{kawa} is not based on
$N_{eff}$ alone,
but rather
on a detailed study of primordial abundances.
This clearly
illustrates
the fact that BBN bounds are highly flavour sensitive,
and
shows that the much simpler energy density bound from CMB
and
large scale
structure leads to a stronger bound on the
reheating
temperature.


\end{document}